%% file: prd.tex
\documentclass[aps,prd,twocolumn,showpacs,groupedaddress,floatfix]{revtex4}
\usepackage{graphicx}  % needed for figures
\usepackage{dcolumn}   % needed for some tables
\usepackage{bm}        % for math
\usepackage{amssymb}   % for math
\usepackage{multirow}
\usepackage{verbatim}
\usepackage{array}
\def\met{{\mbox{$E\kern-0.57em\raise0.19ex\hbox{/}_{T}$}}}

\def\ifb{fb$^{-1}$}
\def\pp{$p\bar{p}$}
\def\tt{$t\bar{t}$}
\def\wjets{$W+$jets}
\def\zjets{$Z+$jets}
\def\tevE{$\sqrt{s}=1.96$~TeV}

\def\pythia{\textsc{Pythia}}
\def\hzw{\textsc{HZW}}

\def\mida{$+2/-3$}
\def\midb{$+0/-5$}

\begin{document}

% the following line is for submission, including submission to the arXiv!!
%\hspace{5.2in} \mbox{Fermilab-Pub-09/376-E}
\title{Measurement of trilinear gauge boson couplings from \\{\boldmath$WW+WZ\rightarrow \ell\nu jj$} events in {\boldmath$p\bar{p}$} collisions at {\boldmath$\sqrt{s}=1.96$} TeV}

\input list_of_authors_r2.tex  % input Dzero author list

\date{July 24, 2009}

\begin{abstract}
  We present a direct measurement of trilinear gauge boson couplings
  at $\gamma WW$ and $ZWW$ vertices in $WW$ and $WZ$ events produced
  in \pp\ collisions at \tevE.  We consider events with one electron
  or muon, missing transverse energy, and at least two jets.  The 
  data were collected using the D0 detector and correspond to
  1.1~\ifb\ of integrated luminosity.  Considering two different
  relations between the couplings at the $\gamma WW $ and $ZWW$
  vertices, we measure these couplings at 68\% C.L. to be
  $\kappa_{\gamma}=1.07^{+0.26}_{-0.29}$, $\lambda
  =0.00^{+0.06}_{-0.06}$, and $g_{1}^{Z}=1.04^{+0.09}_{-0.09}$ in a
  scenario respecting $SU(2)_L\otimes U(1)_Y$ gauge symmetry and 
  $\kappa =1.04^{+0.11}_{-0.11} $ and $\lambda =0.00^{+0.06}_{-0.06}$ 
  in an ``equal couplings'' scenario.
\end{abstract}


\maketitle 

\section{Introduction}
  \label{sec:intro}

  \noindent A primary motivation for studying diboson physics is that
  the production of two weak bosons and their interactions provide
  tests of the electroweak sector of the standard model (SM) arising
  from the vertices involving trilinear gauge boson couplings
  (TGCs)~\cite{bib:anom-d02}. Any deviation of TGCs from their
  predicted SM values would be an indication for new
  physics~\cite{bib:strong} and could provide
  information on a mechanism for electroweak symmetry breaking (EWSB).

  The TGCs involving the $W$ boson have been previously probed in $WW$, 
  $W\gamma$ and $WZ$ production at the Tevatron \pp\
  Collider~\cite{bib:ww,bib:wz,bib:wgamma,bib:cdf} and $WW$ production
  at the CERN $e^{+}e^{-}$ collider
  (LEP)~\cite{bib:aleph,bib:opal,bib:l3,bib:anom-lep1}, at different
  center-of-mass energies and luminosities but no deviation from the
  SM predictions has been observed.  The LEP experiments benefit from
  the full reconstruction of event kinematics in $e^{+}e^{-}$
  collisions, high signal selection efficiencies and small background
  contamination.  At the Tevatron, despite larger backgrounds and
  limited ability to fully reconstruct event kinematics, larger
  collision energies are probed and $WZ$ production can be used to
  directly probe the $ZWW$ coupling. The study of $WW$ and $WZ$
  production at hadron colliders has focused primarily on the purely
  leptonic final states~\cite{bib:ww,bib:wz,bib:xsec}.  In this paper
  we present a measurement of the $\gamma WW/ZWW$ couplings based on
  the same dataset used to obtain the recent evidence for
  semileptonic decays of $WW/WZ$ boson pairs in hadron
  collisions~\cite{bib:ournote}.
  
\begin{figure}[htb] \begin{centering}
    \begin{tabular}{ccc}
      \multirow{1}{*}[1.0in]{(a)} & \includegraphics[height=1.2in]{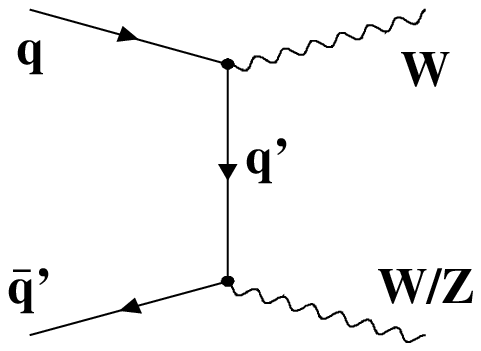}\\
      \multirow{1}{*}[1.0in]{(b)} & \includegraphics[height=1.2in]{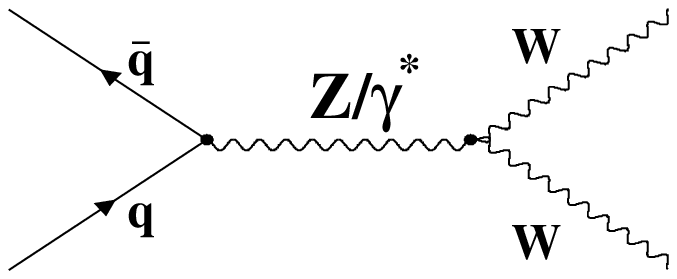}\\
      \multirow{1}{*}[1.0in]{(c)} & \includegraphics[height=1.2in]{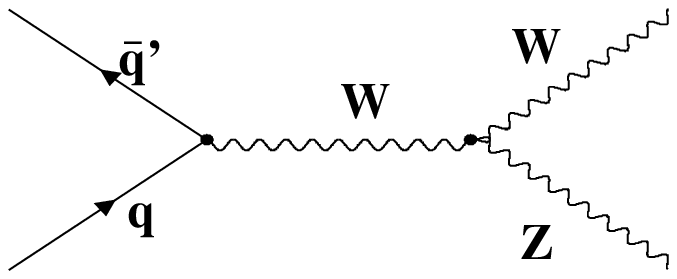}
    \end{tabular}
    \caption{Tree-level Feynman diagrams for the processes of $WW/WZ$ production at the
     Tevatron collider via (a) $t$-channel exchange and (b) and (c) $s$-channel.}
    \label{fig:feynmans} \end{centering}
  \end{figure}

  As shown in the tree-level diagrams of~Fig.~\ref{fig:feynmans}, TGCs
  contribute to $WW/WZ$ production via $s$-channel diagrams.
  Production of $WW$ via the $s$-channel process contains both
  trilinear $\gamma WW$ and $ZWW$ gauge boson vertices.  On the other
  hand, $WZ$ production is sensitive exclusively to the $ZWW$ vertex.

  \section{Phenomenology}
  \label{sec:pheno}

  \noindent
  Unraveling the origins of EWSB and the mass generation mechanism 
  are currently the highest priorities in particle physics.  The 
  SM introduces an effective Higgs potential with an upper limit on the 
  Higgs boson mass of $\simeq$ 1~TeV to prevent tree-level unitarity
  violation~\cite{bib:higgs}.  

  In a Higgs-less scenario or for heavier Higgs boson masses this
  unitarity limit on the Higgs boson mass indicates the mass scale at
  which the SM must be superseded by new physics in order to
  restore unitarity at TeV energies.  In this case, the SM is
  considered to be a low-energy approximation of a general theory.
  Conversely, if a light Higgs boson exists, the SM may nevertheless
  be incomplete and new physics could appear at higher energies.

  The effects of this general theory can be described by an effective
  Lagrangian, ${\mathcal L}_{\rm{eff}}$, describing low-energy
  interactions of the new physics at higher energies in a
  model-independent manner. Expanding in powers of
  (1/$\Lambda_{NP}$)~\cite{bib:lowEnergy}:
  \begin{equation}
  \mathcal{L}_{\rm{eff}} ={\mathcal L}_{\rm{eff}}^{SM}+\sum_{n \ge 1}\sum_{i}\frac{f_{i}}{{\Lambda^{n}_{NP}}}\mathcal{O}_{i}^{(n+4)}
  \label{eq:newphys}
  \end{equation} 
  \noindent where $\mathcal L_{\rm{eff}}^{SM}$ is the
  $SU(2)_L\times{U(1)_Y}$ gauge-invariant SM Lagrangian,
  $\Lambda_{NP}$ is the energy scale of the new physics and $i$ sums
  over all operators ${\mathcal O_{i}}$ of the given energy dimension
  $(n+4)$.  The coefficients $f_{i}$ parametrize all possible
  interactions at low energies.  Effects of the new physics may not be
  directly observable because the scale of the new physics is above
  the energies currently experimentally accessible.  However, there
  could be indirect consequences with measurable effects; for example,
  on gauge boson interactions.

  For the study of gauge boson interactions, the relevant terms
  in~Eq.~\ref{eq:newphys} are those that produce vertices with three
  or four gauge bosons.  The effective Lagrangian,
  $\mathcal{L}_{\rm{eff}}$, that parametrizes the most general Lorentz
  invariant $VWW$ vertices ($V=Z,\gamma$) involving two $W$ bosons can
  be defined as~\cite{bib:hagiwara}
  \begin{equation} \begin{array}{ccl} \frac{{\mathcal
  L}_{\rm{eff}}^{VWW}}{g_{VWW}} & = & i g_{1}^{V}
  (W_{\mu\nu}^{\dag}W^{\mu}V^{\nu} - W_{\mu}^{\dag}V_{\nu}W^{\mu\nu}) \\ & +
  & i{\kappa}_{V}W_{\mu}^{\dag}W_{\nu}V^{\mu\nu} +
  i\frac{\lambda_{V}}{M_{W}^{2}}
  W_{\lambda\mu}^{\dag}W_{\nu}^{\mu}V^{\nu\lambda} \\ & - &
  g_{4}^{V}W_{\mu}^{\dag}W_{\nu}(\partial^{\mu}V^{\nu} +
  \partial^{\nu}V^{\mu}) \\ & + &
  g_{5}^{V}\epsilon^{\mu\nu\lambda\rho}(W_{\mu}^{*}\partial_{\lambda}W_{\nu}-
  \partial_{\lambda}W^{\dag}_{\mu}W_{\nu})V_{\rho} \\ & + &
  i\tilde{\kappa}_{V}W^{\dag}_{\mu}W_{\nu}\tilde{V}^{\mu\nu} +
  i\frac{\tilde{\lambda}_{V}}{M_{W}^{2}}W^{\dag}_{\lambda\mu}W^{\mu}_{\nu}\tilde{V}^{\nu\lambda}
  \label{eq:effLang} \end{array}{} \end{equation} 
  \noindent where $\epsilon_{\mu\nu\lambda\rho}$ is the fully
  antisymmetric $\epsilon$ tensor, $W$ denotes the $W$ boson field,
  $V$ denotes the photon or $Z$ boson field,
  $V_{\mu\nu}=\partial_{\mu}V_{\nu}-\partial_{\nu}V_{\mu}$,
  $W_{\mu\nu}=\partial_{\mu}W_{\nu}-\partial_{\nu}W_{\mu}$,
  $\tilde{V}_{\mu\nu}=1/2(\epsilon_{\mu\nu\lambda\rho}V^{\lambda\rho})$,
  $g_{\gamma WW}=-e$, and $g_{ZWW}=-e\cot\theta_{W}$, where $e$ is the
  electron electric charge, $\theta_{W}$ is the weak mixing angle and
  $M_{W}$ is the $W$ boson mass.  The 14 coupling parameters of
  $VWW$ vertices are grouped according to the symmetry properties of
  their corresponding operators: $C$ (charge conjugation) and $P$
  (parity) conserving ($g_{1}^{V},~{\kappa}_{V}$, and $\lambda_{V}$),
  $C$ and $P$ violating but $CP$ conserving ($g_{5}^{V}$), and $CP$
  violating ($g_{4}^{V}, \tilde{\kappa}_{V}$, and
  $\tilde{\lambda}_{V}$).  In the SM all couplings vanish
  ($g_{5}^{V}=g_{4}^{V}=\tilde{\kappa}_{V}=\tilde{\lambda}_{V}={\lambda}_{V}=0$)
  except $g_{1}^{V}=\kappa_{V}=1$.  The value of $g_{1}^{\gamma}$ is
  fixed by electromagnetic gauge invariance ($g_{1}^{\gamma}=1$) while
  the value of $g_{1}^{Z}$ may differ from its SM value.  Considering
  the $C$ and $P$ conserving couplings only, five couplings remain,
  and their deviations from the SM values are denoted as the anomalous
  TGCs $\Delta{g_{1}^{Z}}=$ $(g_{1}^{Z}-1)$,
  $\Delta\kappa_{\gamma}=(\kappa_{\gamma}-1)$,
  $\Delta\kappa_{Z}=(\kappa_{Z}-1)$, $\lambda_{\gamma}$ and
  $\lambda_{Z}$.

  If non zero anomalous TGCs are introduced in~Eq.~\ref{eq:effLang}, an 
  unphysical increase in the $WW$ and $WZ$ production cross
  sections will result as the center-of-mass energy, $\sqrt{\hat{s}}$, of the
  partonic constituents approaches $\Lambda_{NP}$. Such divergences
  would violate unitarity, but can be controlled by introducing a
  form factor for which the anomalous coupling vanishes as
  $\hat{s}\rightarrow\infty$:
  \begin{equation} \Delta{a(\hat{s})}=
  \frac{\Delta{a_{0}}}{(1+\hat{s}/\Lambda_{NP}^{2})^{n}} \label{eq:alpha}
  \end{equation}
  \noindent where $n=2$ for $\gamma{WW}$ and $ZWW$ couplings, and 
  $a_{0}$ is a low-energy approximation of the coupling $a(\hat{s})$. 
  Thus, the previously described anomalous TGCs scale as $\Delta{a_{0}}$ 
  in~Eq.~\ref{eq:alpha}.  The values of $\Delta{a_{0}}$ (and $a_{0}$) 
  are constrained by requiring the $S$-matrix unitarity condition that 
  bounds the $J=1$ partial-wave amplitude of inelastic vector boson 
  scattering by a constant.  These constants were derived by Baur and 
  Zeppenfeld~\cite{bib:bounds} for each coupling that contributes to 
  reduced helicity amplitudes in $WZ,~\gamma{W}$, or $WW$ production via 
  $s$-channel.  Calculated with $M_{W}=80$~GeV, $M_{Z}=91.1$~GeV 
  and with the dipole form factor as given by~Eq.~\ref{eq:alpha}, the 
  unitarity bounds for $\Delta\kappa_{\gamma},~\Delta\kappa_{Z},~\Delta{g_{1}^{Z}}$ 
  and $\lambda$ TGCs are

  \begin{equation} 
  \begin{array}{ll}
  |\Delta\kappa_{\gamma}^{0}|\leq\frac{n^{n}}{(n-1)^{n-1}}\frac{1.81~\text{TeV}^{2}}{\Lambda_{NP}^{2}},
  ~|\Delta\lambda_{\gamma}^{0}|\leq\frac{n^{n}}{(n-1)^{n-1}}\frac{0.96~\text{TeV}^{2}}{\Lambda_{NP}^{2}} \\
  |\Delta\kappa_{Z}^{0}|\leq\frac{n^{n}}{(n-1)^{n-1}}\frac{0.83~\text{TeV}^{2}}{\Lambda_{NP}^{2}},
  ~|\Delta\lambda_{Z}^{0}|\leq\frac{n^{n}}{(n-1)^{n-1}}\frac{0.52~\text{TeV}^{2}}{\Lambda_{NP}^{2}} \\
  |\Delta{g_{1}^{Z0}}|\leq\frac{n^{n}}{(n-1)^{n-1}}\frac{0.84~\text{TeV}^{2}}{\Lambda_{NP}^{2}} \\
  \label{eq:bounds}
  \end{array}
  \end{equation}

  \noindent For $n=2$ and $\Lambda_{NP}=2$~TeV, the unitarity
  condition sets constraints on the TGCs of
  $|\Delta\kappa_{\gamma}^{0}|\leq{1.81}$,
  $|\Delta\lambda_{\gamma}^{0}|\leq{0.96}$,
  $|\Delta\kappa_{Z}^{0}|\leq{0.83}$,
  $|\Delta\lambda_{Z}^{0}|\leq{0.52}$, and
  $|\Delta{g_{1}^{Z0}}|\leq{0.84}$.  The scale of
  new physics, $\Lambda_{NP}$, was chosen such that the unitarity
  limits are close to, but no tighter than, the coupling limits set by
  data.  Clearly, as $\Lambda_{NP}$ increases the effects on anomalous
  TGCs decrease and their observation requires either more precise
  measurements or higher $\hat{s}$.

  \section{Relations Between Couplings} 
  \label{sec:relAC}

  \noindent The interpretation of the effective
  Lagrangian~[Eq.~\ref{eq:newphys}] depends on the specified symmetry
  and the particle content of the underlying low-energy theory.  In
  general, $\mathcal{L}_{\rm{eff}}$ can be expressed using either the
  linear or nonlinear realization of the $SU(2)_L\times{U(1)_Y}$ 
  symmetry~\cite{bib:nonLin} to prevent unitarity violation, depending 
  on its particle content.  Thus, $\mathcal{L}_{\rm{eff}}$ can be 
  rewritten in a form that includes the operators that describe 
  interactions involving additional gauge bosons, and/or Goldstone bosons, 
  and/or the Higgs field and operators of interest for any new physics 
  effects.  The number of operators can be reduced by considering their 
  detectable contribution to the measured coupling.

  Assuming the existence of a light Higgs boson, the low-energy
  spectrum is augmented by the Higgs doublet field $\phi$, and
  $SU(2)_L$ and $U(1)_Y$ gauge fields.  Because experimental evidence
  is consistent with the existence of an $SU(2)_L\times{U(1)_Y}$ gauge
  symmetry, it is reasonable to require $\mathcal{L}_{\rm{eff}}$ to be
  invariant with respect to this symmetry.  Thus, the second term in
  Eq.~\ref{eq:newphys} consisting of operators up to energy dimension
  six, is also required to have local $SU(2)_L\times{U(1)_Y}$ gauge
  symmetry and the underlying physics is described using a linear
  realization~\cite{bib:linNoNlin} of the $SU(2)_L\times{U(1)_Y}$
  symmetry.  By considering operators that give rise to nonstandard
  $\gamma{WW}$ and $ZWW$ couplings at the tree level,
  $\mathcal{L}_{\rm{eff}}$ can be parametrized in terms of the
  $\alpha_{i}$ parameters~\cite{bib:leppar}.  Those parameters relate
  to the $f_{i}$ parameters of the Lagrangian given
  in~Eq.~\ref{eq:newphys} and to the TGCs in the Lagrangian
  of~Eq.~\ref{eq:effLang} as follows~\cite{bib:hisz}:

  \begin{equation} 
  \begin{array}{lll}
  \Delta\kappa_{\gamma}=(f_{W\phi}+f_{B\phi})\frac{M_{W}^{2}}{2\Lambda_{NP}^{2}} =\alpha_{W\phi}+\alpha_{B\phi} \\
  \Delta{g_{1}^{Z}}= f_{W\phi}\frac{M_{Z}^{2}}{2\Lambda_{NP}^{2}}=
  \Delta\kappa_{Z}+\frac{s_{W}^{2}}{c_{W}^{2}}\Delta\kappa_{\gamma}=\frac{\alpha_{W\phi}}{c_{W}^{2}} \\
  \lambda=\lambda_{\gamma} = \lambda_{Z} = 3g^{2}\frac{M_{W}^{2}}{2\Lambda_{NP}^{2}}f_{WWW}= \alpha_{W}
  \label{eq:relLEP}
  \end{array}
  \end{equation} 
  where $g$ is the $SU(2)_{L}$ gauge coupling constant
  ($g=e/{\text{sin}}\theta_{W}$), $c_{W}=\cos\theta_{W}$,
  $s_{W}=\sin\theta_{W}$, and indices $W\phi$ ($B\phi$) and $W$ refer
  to operators that describe the interactions between the $W$ ($B$)
  gauge boson field and the Higgs field $\phi$, and the gauge boson
  field interactions, respectively.  The relations
  in~Eq.~\ref{eq:relLEP} give the expected order of magnitude for TGCs
  to be $\mathcal{O}(M_{W}^{2}/\Lambda_{NP}^{2})$.  Thus, for
  $\Lambda_{NP}\approx{2}$~TeV, the expected order of magnitude for
  $\Delta\kappa_{\gamma}$, $\Delta{g^{Z}_{1}}$, and $\lambda$ is
  $\mathcal{O}(10^{-3})$.
  This gauge-invariant parametrization, also used at LEP, gives the
  following relations between the $\Delta\kappa_{\gamma}$,
  $\Delta{g^{Z}_{1}}$ and $\lambda$ couplings:

  \begin{equation} \begin{array}{ccl}
  \Delta\kappa_{Z} = \Delta g^{Z}_{1}-\Delta\kappa_{\gamma}\cdot
  \tan^{2}\theta_{W}~\text{and}~\lambda \equiv \lambda_{Z} = \lambda_{\gamma} \label{eq:lep} \end{array}{} \end{equation} 
  \noindent Hereafter we will refer to this relationship as the ``LEP
  parametrization'' [or SU(2)xU(1) respecting scenario] with three
  different parameters: $\Delta\kappa_{\gamma},~\lambda$ and
  $\Delta{g^{Z}_{1}}$.  The coupling $\Delta\kappa_{Z}$ can be
  expressed via the relation given by~Eq.~\ref{eq:lep}.

  A second interpretive scenario, referred to as the equal
  couplings (or $ZWW$$=$$\gamma WW$) scenario~\cite{bib:anom-d02}, specifies the $\gamma{WW}$
  and $ZWW$ couplings to be equal.  This is also relevant for studying
  interference effects between the photon and $Z$-exchange diagrams in
  $WW$ production (see~Fig.~\ref{fig:feynmans}).  In this case,
  electromagnetic gauge invariance forbids any deviation of $g^{\gamma}_{1}$
  from its SM value ($\Delta g^{Z}_{1}=\Delta g^{\gamma}_{1}=0$ ) and
  the relations between the couplings become
  \begin{equation} \begin{array}{ccl} 
\Delta\kappa \equiv \Delta\kappa_{Z} =
  \Delta\kappa_{\gamma} & \text{and} & \lambda \equiv \lambda_{Z} = \lambda_{\gamma}
  \label{eq:equal} \end{array}{} \end{equation} 

  As already stated, for $WW$ and $WZ$ production the anomalous
  couplings contribute to the total cross section via the $s$-channel
  diagram.  Anomalous couplings enter the differential production cross 
  sections through different helicity amplitudes that depend on $\hat{s}$.  
  The coupling $\lambda$ primarily affects transversely polarized
  gauge bosons, which is the main contribution to the total cross
  section.  Consequently, for a given $\hat{s}$, the sensitivity to
  the coupling $\lambda$ is higher than to $\kappa$ because $\lambda$
  is multiplied by $\hat{s}$ in dominating amplitudes for $WW$ and
  $WZ$ production. Different sensitivity to the $\kappa$ couplings
  is expected due to the choice of scenario: the sensitivity to the
  $\kappa$ coupling in the equal couplings scenario is higher than
  in the LEP parametrization scenario simply because of the
  different relations between~Eq.~\ref{eq:lep} and~Eq.~\ref{eq:equal}.

  \section{D0 Detector}
  \label{sec:detector}

  \noindent The analyzed data were produced in \pp\ collisions at
  \tevE\ by the Tevatron collider at Fermilab and collected by the D0
  detector~\cite{bib:d0det} during 2002 - 2006.  They correspond to
  $1.07\pm{0.07}$ fb$^{-1}$ of integrated luminosity for each of the two lepton
  channels ($e\nu q\bar q$ and $\mu\nu q\bar q$).

  The D0 detector is a general purpose collider detector consisting of
  a central tracking system, a calorimeter system, and an outer muon
  system.  The central tracking system consists of a silicon
  microstrip tracker and a central fiber tracker, both located within
  a 2~T superconducting solenoidal magnet, with
  designs optimized for tracking and vertexing at
  pseudorapidities~\cite{bib:footnote} $|\eta|<3$ and $|\eta|<2.5$,
  respectively.  A liquid-argon and uranium calorimeter has a central
  section covering pseudorapidities $|\eta|$ up to $\approx 1.1$, and
  two end calorimeters that extend coverage to $|\eta|\approx 4.2$,
  with all three housed in separate cryostats~\cite{bib:run1det}. An
  outer muon system, covering $|\eta|<2$, consists of a layer of tracking
  detectors and scintillation trigger counters in front of 1.8~T iron
  toroids, followed by two similar layers after the
  toroids~\cite{bib:run2muon}.

  Jets at D0 are reconstructed using the Run II cone
  algorithm~\cite{bib:jetalg} with cone radius $R=\sqrt{(\Delta
  y)^{2}+(\Delta\phi)^{2}}=0.5$; where $y$ is the rapidity.  Jet energies
  are corrected to the particle level. The jet energy resolution for
  data, defined as $\sigma_{p_{T}}/p_{T}$, ranges from $\sim{15\%-25}\%$ 
  for jets with $p_{T}=20$~GeV to $\sim{7\%-12}\%$ for jets with $p_{T}=300$~GeV, 
  depending on the rapidity of the jet.
 
  The D0 detector uses a three-level trigger system for quickly
  filtering events from a rate of 1.7 MHz down to around 100 Hz that
  are stored for analysis.  Events analyzed in the electron channel
  had to pass a trigger based on a single electron or electron+jet(s)
  requirement, resulting in an efficiency of $98^{+2}_{-3}\%$.  The
  triggers based on specific single muon and muon+jet(s) requirements
  are about 70\% efficient.  Thus, all available triggers were used
  for the muon channel to achieve higher efficiency.  We select all
  events that satisfy our kinematic selection requirements with no
  specific trigger requirement.  The efficiency in this kinematic
  region is very nearly 100\%.  To estimate and account for possible
  biases on the shape of kinematic distributions, we compare data
  selected with the inclusive triggers to data selected with triggers
  based on a single muon.  In the kinematic region of interest, the
  inclusive trigger is estimated to have a shape uncertainty of less
  than 5\% and a normalization uncertainty of 2\%.

  \section{Event Selection and Cross Section Measurement}
  \label{sec:selection}

  \noindent The analysis presented here builds upon a previous
  publication in which we reported the first evidence of $WW/WZ$
  production with semileptonic final states at a hadron
  collider~\cite{bib:ournote}.  Such events have two energetic jets
  from the hadronic decay of either a $W$ or $Z$ boson as well as an
  energetic charged lepton and significant missing transverse energy
  (indicating a neutrino) from the leptonic decay of the associated
  $W$ boson.  
%% What is the trigger selection?  
  Therefore, at the analysis level, we selected events with a reconstructed electron or muon
  with transverse momentum $p_T\ge 20$~GeV and pseudorapidity
  $|\eta|\le 1.1\ (2.0)$ for electrons (muons), a missing transverse
  energy of $\met\ge 20$~GeV, and at least two jets with $p_T\ge
  20$~GeV and $|\eta|\le 2.5$.  The jet of highest $p_T$ was required
  to satisfy $p_T\ge 30$~GeV.  To reduce background from processes
  that do not contain $W$$\rightarrow \ell\nu$, we required the
  transverse mass~\cite{bib:trans} from the lepton and $\met$ to be
  $M_T^{\ell\nu}\ge 35$~GeV.  The multijet background, for which a jet
  is misidentified as a lepton, was estimated using independent data
  samples.

  Signal ($WW$ and $WZ$) and background (\wjets, \zjets, \tt\ and
  single top quark) processes were modeled using Monte Carlo (MC) 
  simulation.  All MC samples were normalized using next-to-leading-order 
  (NLO) or next-to-next-to-leading-order predictions for SM cross 
  sections, except the dominant background \wjets, which was scaled 
  to match the data as described below.

  In the previously published cross section measurement
  analysis~\cite{bib:ournote}, the signal and backgrounds were further
  separated using a multivariate classifier to combine information
  from several kinematic variables.  The multivariate classifier
  chosen was a random forest (RF) classifier~\cite{bib:SPR1,bib:SPR2}.
  Thirteen well-modeled kinematic variables that demonstrated a
  difference in probability density between signal and at least one of
  the backgrounds were used as inputs to the RF.  The effects of
  systematic uncertainties on the normalization and on the shape of
  the RF distributions were evaluated for signal and backgrounds.

\begin{figure}[htb] \begin{centering}
  \includegraphics[width=8.5cm]{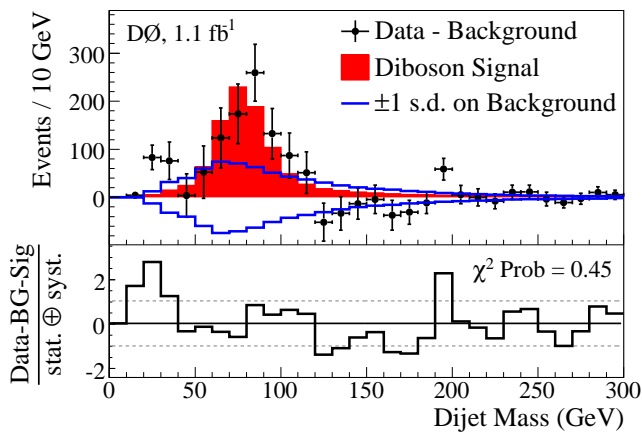}
  \caption{A comparison of the extracted signal (filled histogram) to
    background-subtracted data (points), along with the $\pm1$
    standard deviation (s.d.) systematic uncertainty on the
    background. The residual distance between the data points and the
    extracted signal, divided by the total uncertainty, is given
    at the bottom.}  \label{fig:dijetmass} \end{centering}
  \end{figure}

  The signal cross section was determined from a fit of signal and
  background RF output distributions to the data by minimizing a
  Poisson $\chi^2$ function ({\it i.e.}, a negative log likelihood)
  with respect to variations of the systematic
  uncertainties~\cite{bib:pflh}, assuming SM $\gamma WW$ and $ZWW$
  couplings.  The fit simultaneously varied the $WW/WZ$ and \wjets\
  contributions, thereby also determining the normalization factor for
  the \wjets\ MC sample.  The measured yields for signal and each
  background are given in Table~\ref{tab:yields} and the dijet mass
  peak extracted from data compared to the $WW/WZ$ MC prediction is
  shown in~Fig.~\ref{fig:dijetmass}.  The combined fit of both
  channels to the RF output resulted in a measured cross section of
  20.2~$\pm$~2.5(stat)~$\pm$~3.6(syst)~$\pm$~1.2(lumi)~pb, which is
  consistent with the NLO SM predicted cross section of
  $\sigma(WW+WZ)=16.1\pm 0.9$~pb~\cite{bib:Campbell}.
  
  \begin{table}[htb]
    \caption{Measured number of events for signal and each background
    after the combined fit (with total uncertainties determined from
    the fit) and the number observed in data.}
    \label{tab:yields}
    \begin{ruledtabular}
      \begin{tabular}{l @{\extracolsep{\fill}} r @{$\ \pm\ $\extracolsep{0cm}} l @{\extracolsep{\fill}} r @{$\ \pm\ $\extracolsep{0cm}} l}
	           & \multicolumn{2}{c}{$e\nu q\bar{q}$ channel} & \multicolumn{2}{c}{$\mu\nu q\bar{q}$ channel} \\
	\hline	   
	Diboson signal        &   436  &  36  &   527  &  43 \\
	$W$+jets              & 10100  & 500  & 11910  & 590 \\
	$Z$+jets              &   387  &  61  &  1180  & 180 \\
	\tt\ + single top  &   436  &  57  &   426  &  54 \\
	Multijet              &  1100  & 200  &   328  &  83 \\
        \hline
	Total predicted       & 12460  & 550  &  14370 & 620 \\
	Data                  & \multicolumn{2}{c}{12473} & \multicolumn{2}{c}{14392} \\
      \end{tabular}
    \end{ruledtabular}
  \end{table}

  \section{Sensitivity to Anomalous Couplings} \label{sec:sensy}

  For TGCs analysis we use the same selection and set limits on
  anomalous TGCs using a kinematic variable that is highly sensitive
  to the effects of deviations of $\Delta\kappa,~\lambda$, and
  $\Delta{g_{1}^{Z}}$.  Because TGCs introduce terms in the Lagrangian
  that are proportional to the momentum of the weak boson, the
  differential and the total cross sections will deviate from the SM
  prediction in the presence of anomalous couplings.  This behavior is
  also expected at large production angles of a weak boson.  Thus, the
  weak boson transverse momentum spectrum, $p_{T}$, is sensitive to
  anomalous couplings and can show a significant enhancement at high
  values of $p_{T}$.

  The predicted $WW$ and $WZ$ production cross sections in the
  presence of anomalous TGCs are generated with the leading order (LO)
  MC generator of Hagiwara, Zeppenfeld, and
  Woodside~(\hzw)~\cite{bib:anom-d02} with
  \textsc{CTEQ5L}~\cite{bib:CTEQ} parton distribution functions
  (PDFs). For example, the predicted ``anomalous'' cross sections
  relative to the SM value given by the~\hzw\ generator are shown
  in~Fig.~\ref{fig:x-secs} as a function of anomalous couplings. For
  this figure we vary only the $\Delta\kappa$ coupling with the
  constraint between $\Delta\kappa_{\gamma}$ and $\Delta\kappa_{Z}$ as
  given by~Eq.~\ref{eq:lep}.  The couplings $\lambda$ and
  $\Delta{g_{1}^{Z}}$ are fixed to their SM values ({\it i.e.},
  $\lambda=\Delta{g_{1}^{Z}}=$ 0).  The effects of anomalous couplings
  on two $WW$ kinematic distributions ($p_{T}$ and rapidity of the 
  $q\bar{q}$ system) for the
  LEP parametrization are shown in~Fig.~\ref{fig:kine}.  Here
  again, we vary only one coupling at a time ($\Delta\kappa,~\lambda$
  or $\Delta{g_{1}^{Z}}$) according to~Eq.~\ref{eq:lep} and leave the
  others fixed to their SM values.  Finally, we choose the
  $p_{T}^{q\bar{q}}$ ({\it i.e.}, reconstructed dijet $p_{T}$)
  distribution to be our kinematic variable to probe anomalous
  couplings in data.  Results are interpreted in two different
  scenarios: LEP parametrization and equal couplings, both
  with $\Lambda_{NP}=$ 2~TeV.

  \section{Reweighting Method}
  \label{sec:method}
  
  \noindent The \pythia~\cite{bib:PYTHIA} LO MC generator with
  \textsc{CTEQ6L1} PDFs was used to simulate a sample of $WW$ and $WZ$
  events at LO.  We use the \textsc{mc@nlo} MC
  generator~\cite{bib:mc@nlo} with \textsc{CTEQ6M} PDFs to correct the
  event kinematics for higher order QCD effects by reweighting the
  differential distributions of $p_T(WV)$ and $\Delta R(W,V)$ produced
  by \pythia\ to match those produced via \textsc{mc@nlo}.  We
  simulate the LO effects of anomalous couplings on the $p_{T}$
  distribution by reweighting the SM predictions for $WW$ and $WZ$
  production from \pythia\ to include the contribution from the
  presence of anomalous couplings.  The anomalous coupling
  contribution to the normalization and to the shape of
  $p_{T}^{q\bar{q}}$ distribution relative to the SM is predicted by
  the \hzw\ LO MC generator.
  
  The reweighting method uses the matrix element values given by the
  generator to predict an event rate in the presence of anomalous
  couplings.  More precisely, an event rate ($R$) is assigned representing 
  the ratio of the differential cross section with anomalous couplings 
  to the SM differential cross section.  Because
  the \hzw\ generator does not recalculate matrix element values, we
  use high statistics samples to estimate the weight as a function of
  different anomalous couplings.  Thus, we consider our approach to be
  a close approximation of an exact reweighting method.  

  The basis of the reweighting method is that, in general, the
  equation of the differential cross section, which has a quadratic
  dependence on the anomalous couplings, can be written as
  \begin{equation}
  \begin{array}{ccl} d\sigma & = & {\textit{const}}\cdot|\mathcal
  M|^{2}dX \\ & = & {\textit{const}}\cdot|\mathcal
  M|_{SM}^{2}\frac{|{\mathcal M}|^{2}}{|{\mathcal M}|_{SM}^{2}}dX \\ &
  = & {\textit{const}}\cdot|\mathcal
  M|^{2}_{SM}[1+A(X)\Delta\kappa+B(X)\Delta\kappa^{2}\\ & + &
  C(X)\lambda + D(X)\lambda^{2}+E(X)\Delta\kappa\lambda
  + ...]dX \\ & = & d\sigma_{SM}\cdot R(X;\Delta\kappa,\lambda,
  ...)  \label{eq:expan} \end{array}{} \end{equation} 
  where $d\sigma$ is the differential cross section that includes the
  contribution from the anomalous couplings, $d\sigma_{SM}$ is the SM
  differential cross section, $X$ is a kinematic distribution
  sensitive to the anomalous couplings and $A(X),~B(X),~C(X),~D(X)$,
  and $E(X)$ are reweighting coefficients dependent on $X$.
  
  In the LEP parametrization,~Eq.~\ref{eq:expan} is parametrized
  with the three couplings $\Delta\kappa_{\gamma},~\lambda$, and
  $\Delta{g_{1}^{Z}}$ and nine reweighting coefficients, $A(X)-I(X)$.
  Thus, the weight $R$ in the LEP parametrization scenario is
  defined as
  
  \begin{equation}
  \begin{array}{ccl}
  &R&(X;\Delta\kappa,\lambda,\Delta g_{1}) = 1+A(X)\Delta\kappa \\
  &+& B(X)(\Delta\kappa)^{2}+C(X)\lambda+D(X)\lambda^{2} \\
  &+& E(X)\Delta g_{1}+F(X)(\Delta g_{1})^{2}+G(X)\Delta\kappa\lambda \\
  &+& H(X)\Delta\kappa \Delta g_{1}+I(X)\lambda\Delta g_{1} \\
  \label{eq:expan1} \end{array}{} \end{equation}
  with $\Delta\kappa=\Delta\kappa_{\gamma}$,
  $\lambda=\lambda_{\gamma}=\lambda_{Z}$, and
  $\Delta{g_{1}}=\Delta{g_{1}^{Z}}$.

  In the equal couplings scenario,~Eq.~\ref{eq:expan} is
  parametrized with the two couplings $\Delta\kappa$ and $\lambda$
  and five reweighting coefficients, $A(X)-E(X)$.  In this case the
  weight is defined as
  \begin{equation} \begin{array}{ccl} &R&(X;\Delta\kappa,\lambda) =
  1+A(X)\Delta\kappa+B(X)\Delta\kappa^{2} \\
  &+&C(X)\lambda+D(X)\lambda^{2}+E(X)\Delta\kappa\lambda
  \label{eq:expan2} \end{array}{} \end{equation}
  with $\Delta\kappa=\Delta\kappa_{\gamma}=\Delta\kappa_{Z}$ and 
  $\lambda=\lambda_{\gamma}=\lambda_{Z}$.  

  The kinematic variable $X$ is chosen to be the $p_{T}$ of the $q\bar{q}$ 
  system, which is highly sensitive to anomalous couplings, as demonstrated 
  in~Fig.~\ref{fig:kine}.  Depending on the number of reweighting coefficients, 
  a system of the same number of equations allows us to calculate their values 
  for each event.  Applied on the SM distribution of $X$ for any combination 
  of anomalous couplings, the distribution of $X$ weighted by $R$ corresponds 
  to the kinematic distribution in the presence of the given non-SM TGC.  

  To calculate reweighting coefficients in the LEP parametrization scenario, 
  we generate nine different functions,
  $F_{i}$ ($i=1-9$), fitting the shape of the $p_{T}^{q\bar{q}}$
  distributions in the presence of anomalous couplings.  The values of
  anomalous TGCs are chosen to deviate $\pm$ 0.5 relative to the SM as
  shown in~Table~\ref{tab:set1}.  We calculate nine weights $R_{i}$
  normalizing the functions $F_{i}$ with the cross sections given by
  the \hzw\ generator.
  \begin{table}[h] 
\begin{ruledtabular}
  \begin{center}
  \caption{The values of $\Delta\kappa_{\gamma},~\lambda$ and $\Delta{g_{1}^{Z}}$
  used to calculate the reweighting coefficients $A(X)-I(X)$ in the
  LEP parametrization scenario.}  
  \label{tab:set1} 
  \begin{tabular}{cccccccccc} 
  & $F_1$& $F_2$& $F_3$&  $F_4$& $F_5$& $F_6$& $F_7$& $F_8$& $F_9$ \\ 
  \hline
  $\Delta\kappa_{\gamma}$ & 0 & 0 & +0.5 & -0.5 & 0 & 0 & +0.5 & +0.5  & 0 \\
  $\lambda$ & +0.5 & -0.5& 0 & 0 & 0 & 0 & +0.5 & 0 & +0.5 \\
  $\Delta{g_{1}^{Z}}$ & 0 & 0 & 0 & 0 & +0.5 & -0.5 & 0 & +0.5 & +0.5 \\
  \end{tabular} 
  \end{center} 
\end{ruledtabular}
  \end{table}

  To verify the derived reweighting parameters, we calculated the
  weight $R$ for different $\Delta\kappa,~\lambda$, and/or
  $\Delta{g_{1}^{Z}}$ values, applied the reweighting coefficients and
  compared reweighted $p_{T}^{q\bar{q}}$ shapes to those predicted by
  the generator.  Discrepancies in the $p_{T}^{q\bar{q}}$ shape of
  less than 5\% and in normalization of less than 0.1\% from those
  predicted by the generator represent reasonable agreement.

  When measuring TGCs in the LEP parametrization, we vary two
  of the three couplings at a time, leaving the third coupling fixed
  to its SM value.  This gives the three two-parameter combinations
  ($\Delta\kappa,~\lambda$), ($\Delta\kappa,~\Delta g_{1}^{Z}$), and
  ($\lambda,~\Delta g_{1}^{Z}$).  For the equal couplings scenario
  there is only the ($\Delta\kappa,~\lambda$) combination.  In each case, 
  the two couplings being evaluated are each varied between -1 and +1 
  in steps of 0.01.  For a given pair of anomalous coupling values, 
  each event in a reconstructed dijet $p_{T}$ bin is weighted by the 
  appropriate weight $R$ and all the weights are summed in that bin.  
  The observed limits are determined from a fit of background and 
  reweighted signal MC distributions for different anomalous couplings 
  contributions to the observed data using the dijet $p_{T}$ distribution 
  of candidate events.

  \section{Systematic Uncertainties}
  \label{sec:syst}

  \noindent
  We consider two general types of systematic uncertainties.  Uncertainties 
  of the first class (\textsc{type I}) are related to the overall normalization 
  and efficiencies of the various contributing physical processes.  The largest 
  contributing~\textsc{type I} uncertainties are those related to the accuracy 
  of the theoretical cross section used to normalize the background processes.  
  These uncertainties are considered to arise from Gaussian parent distributions.
  The second class (\textsc{type II}) consists of uncertainties that, when 
  propagated through the analysis selection, impact the shape of the dijet 
  $p_{T}$ distribution.  The dependence of the dijet $p_{T}$ distribution on 
  these uncertainties is determined by varying each parameter by its
  associated uncertainty ($\pm1$ s.d.) and reevaluating the shape
  of the dijet $p_{T}$ distribution.  The resulting shape dependence
  is considered to arise from a Gaussian parent distribution.
  Although \textsc{type II} uncertainties may also impact efficiencies
  or normalization, any uncertainty shown to impact the shape of the
  dijet $p_{T}$ distribution is treated as \textsc{type II}.  Both
  types of systematic uncertainty are assumed to be 100\% correlated
  amongst backgrounds and signals.  All sources of systematic
  uncertainty are assumed to be mutually independent, and no
  intercorrelation is propagated.  A list of the systematic
  uncertainties used in this analysis can be found in
  Table~\ref{tab:systEMMU}.

  \begin{table*}[htb] 

   \caption{Systematic uncertainties in percent for Monte Carlo
    simulations and multijet estimates. Uncertainties are identical
    for both lepton channels except where otherwise indicated. The
    nature of the uncertainty, {\it i.e.}, whether it refers to a
    normalization uncertainty (\textsc{type I}) or a shape dependence
    (\textsc{type II}), is also provided. The values for
    uncertainties with a shape dependence correspond to the maximum
    amplitude of shape fluctuations in the dijet $p_{T}$ distribution
    (0~GeV $\le p_{T}\le$ 300~GeV) after $\pm1$ s.d. parameter
    changes.  However, the full shape dependence is included in the
    calculations.}
      
    \label{tab:systEMMU}
    \begin{ruledtabular}
      \begin{tabular}{l @{\extracolsep{\fill}} 
      r @{\ \ \extracolsep{\fill}} l @{\extracolsep{\fill}\ \ } 
      r @{\ \ \extracolsep{\fill}} l @{\extracolsep{\fill}\ \ } 
      r @{\ \ \extracolsep{\fill}} l @{\extracolsep{\fill}\ \ } 
      r @{\ \ \extracolsep{\fill}} l @{\extracolsep{\fill}\ \ } 
      r @{\ \ \extracolsep{\fill}} l @{\extracolsep{\fill}\ \ } 
      c
      c}
	\multicolumn{1}{c}{Source of systematic}
        & \multicolumn{2}{c}{Diboson signal}
	& \multicolumn{2}{c}{$W$+jets}
	& \multicolumn{2}{c}{$Z$+jets}
	& \multicolumn{2}{c}{Top}
	& \multicolumn{2}{c}{Multijet}
	& \multicolumn{1}{c}{\multirow{2}{*}{Type}}\\
	\multicolumn{1}{c}{uncertainty}
        & \multicolumn{2}{c}{[\%]}
	& \multicolumn{2}{c}{[\%]}
	& \multicolumn{2}{c}{[\%]}
	& \multicolumn{2}{c}{[\%]}
	& \multicolumn{2}{c}{[\%]}
	&
	&\\

	\hline 

	Trigger efficiency, electron channel\footnote[1]{ Lepton
	  efficiencies depend on kinematics; however, their fractional
	  uncertainties are much less kinematically dependent and have a
	  negligible effect on the shape of the dijet $p_T$
	  distribution.}

                                                             &&   \mida  &&  \mida   &&  \mida   &&  \mida   &&         & I\\
        Trigger efficiency, muon channel                     &&   \midb  &&  \midb   &&  \midb   &&  \midb   &&         & II\\
	Lepton identification\footnotemark[1] &&  $\pm$4  &&  $\pm$4  &&  $\pm$4  &&  $\pm$4  &&         & I\\
	Jet identification                                   &&  $\pm$1  &&  $\pm$1  &&  $\pm$1  &&  $\pm<$1  &&        & II\\
	Jet energy scale                                     &&  $\pm$4  &&  $\pm$7  &&  $\pm$5  &&  $\pm$5  &&         & II\\
	Jet energy resolution                                &&  $\pm$3  &&  $\pm$4  &&  $\pm$4  &&  $\pm$4  &&         & I\\
	Luminosity                                           && $\pm$6.1 && $\pm$6.1 && $\pm$6.1 && $\pm$6.1 &&         & I\\
	Cross section (including PDF uncertainties)          &&          && $\pm$20  &&  $\pm$6  &&  $\pm$10 &&         & I\\
	Multijet normalization, electron channel             &&          &&          &&          &&          && $\pm$20 & I\\
	Multijet normalization, muon channel                 &&          &&          &&          &&          && $\pm$30 & I\\
	Multijet shape, electron channel                     &&          &&          &&          &&          && $\pm$7  & II\\
	Multijet shape, muon channel                         &&          &&          &&          &&          && $\pm$10 & II\\
	Diboson signal NLO/LO shape                          &&  $\pm$10 &&          &&          &&          &&         & II\\
        Diboson signal reweighting shape                     &&  $\pm$5  &&          &&          &&          &&         & II\\
        Parton distribution function (acceptance only)       &&  $\pm$1  &&  $\pm$3  &&  $\pm$2  &&  $\pm$2  &&         & II\\ 
	{\sc alpgen} $\eta$ and $\Delta R$ corrections       &&          &&  $\pm$1  &&  $\pm$1  &&          &&         & II\\
	Renormalization and factorization scale              &&          &&  $\pm$1  &&  $\pm$1  &&          &&         & II\\ 
	{\sc alpgen} parton-jet matching parameters          &&          &&  $\pm$1  &&  $\pm$1  &&          &&         & II\\ 
      \end{tabular}
    \end{ruledtabular}
  \end{table*}

  \section{Anomalous Coupling Limits}
  \label{sec:limits}

  \noindent
  The fit utilizes the \textsc{Minuit}~\cite{bib:minuit} software package to 
  minimize a Poisson $\chi^2$ with respect to variations to the systematic 
  uncertainties~\cite{bib:pflh}.  The $\chi^2$ function used is
  \begin{eqnarray*}
  \chi^2 &=& -2\ln \left (\prod_{i=1}^{{N_b}}\frac{\mathcal{L}^{P}(d_i ; m_i ( \vec{R} ) ) }{\mathcal{L}^{P}(d_i; d_i)} \prod_{k=1}^{{N_s}}\frac{\mathcal{L}^{G}(R_k \sigma_k; 0, \sigma_k)}{\mathcal{L}^{G}(0;0, \sigma_k)}\right ) \\
  &=&  2\sum_{i=1}^{N_b} m_i(\vec{R}) - d_i - d_i \ln\left(\frac{m_i (\vec{R})}{ d_i}\right) + \sum_{k=1}^{N_s} R_k^2~,
  \label{eqn:dchi2}
  \end{eqnarray*}
  \noindent in which the indices $i$ and $k$ run over the number of
  histogram bins ($N_b$) and the number of systematic uncertainties
  ($N_s$), respectively.  In this function ${\mathcal
  L}^P(\alpha;\beta)$ is the Poisson probability for $\alpha$ events
  with a mean of $\beta$ events; ${\mathcal L}^G(x;\mu,\sigma)$ is the
  Gaussian probability for $x$ events in a distribution with a mean
  value of $\mu$ and a variance $\sigma^2$; $R_k$ is a dimensionless
  parameter describing departures in nuisance parameters in units of
  the associated systematic uncertainty $\sigma_k$; $d_i$ is the
  number of data events in bin $i$; and $m_i(\vec{R})$ is the number
  of predicted events in bin $i$~\cite{bib:pflh}.

  Systematics are treated as Gaussian-distributed uncertainties on the
  expected numbers of signal and background events.  The individual
  background contributions are fitted to the data by minimizing this
  $\chi^2$ function over the individual systematic
  uncertainties~\cite{bib:pflh}.  The fit computes the optimal central
  values for the systematic uncertainties, while accounting for
  departures from the nominal predictions by including a term in the
  $\chi^2$ function that sums the squared deviation of each systematic
  in units normalized by its $\pm1$ s.d. uncertainties.
 
  Figure~\ref{fig:subtracted} shows the dijet $p_{T}$ distributions in
  the combined electron and muon channels after the fit.  The value of
  $\chi^2$ is measured between data and MC dijet $p_{T}$ distributions
  as the signal MC is varied in the presence of anomalous couplings.
  The $\Delta\chi^2$ values of 1 and 3.84 from the minimum $\chi^2$ in
  the parameter space, for which all other anomalous couplings are
  zero, represent the 68\% confidence level (C.L.) and 95\%
  C.L. limits, respectively.  For the LEP parametrization, the
  most probable coupling values as measured in data with associated
  uncertainties at 68\% C.L. are $\kappa_{\gamma}=1.07^{+0.26}_{-0.29}$,
  $\lambda=0.00^{+0.06}_{-0.06}$, and $g_{1}^{Z}=1.04^{+0.09}_{-0.09}$.
  For the equal couplings scenario the most probable coupling
  values as measured in data with associated uncertainties at 68\%
  C.L. are $\kappa=1.04^{+0.11}_{-0.11}$ and
  $\lambda=0.00^{+0.06}_{-0.06}$.  The observed 95\% C.L. limits
  estimated from the single parameter fit are -0.44
  $<\Delta\kappa_{\gamma} <$ 0.55, -0.10 $< \lambda <$ 0.11, and -0.12
  $< \Delta g_{1}^{Z} <$ 0.20 for the LEP parametrization or
  -0.16 $< \Delta\kappa <$ 0.23 and -0.11 $< \lambda <$ 0.11 for the
  equal couplings scenario (Table~\ref{tab:results}).

  The observed 68\% C.L. and 95\% C.L. limits in two-parameter space are 
  shown in Figs.~\ref{fig:limit1} and~\ref{fig:limit2} as a function of 
  anomalous couplings along with the most probable values of 
  $\Delta\kappa,~\lambda$, and $\Delta{g_{1}^{Z}}$.
 \begin{table*}[htb]

    \caption{The most probable values with total uncertainties
     (statistical and systematic) at 68\% C.L. for $\kappa_{\gamma}$,
     $\lambda$, and $g_{1}^{Z}$ along with observed 95\%
     C.L. one-parameter limits on $\Delta\kappa_{\gamma}$, $\lambda$,
     and $\Delta g_{1}^{Z}$ measured in 1.1~fb$^{-1}$ of
     $WW/WZ\rightarrow \ell \nu jj$ events with $\Lambda_{NP}=$ 2~TeV.}

    \label{tab:results}
    \begin{ruledtabular}
      \begin{tabular}{l c c c}
68\% C.L. & $\kappa_{\gamma}$ & $\lambda=\lambda_{\gamma}=\lambda_{Z}$ & $g_{1}^{Z}$ \\ \hline
 & & & \\
LEP parametrization & $\kappa_{\gamma}=1.07^{+0.26}_{-0.29}$ & $\lambda = 0.00^{+0.06}_{-0.06}$ & $g_{1}^{Z}=1.04^{+0.09}_{-0.09}$\\
Equal couplings & $\kappa_{\gamma}=\kappa_{Z} = 1.04^{+0.11}_{-0.11}$ & $\lambda = 0.00^{+0.06}_{-0.06}$ &  \\ \hline
95\% C.L. & $\Delta\kappa_{\gamma}$ & $\lambda = \lambda_{\gamma}=\lambda_{Z}$ & $\Delta{g_{1}^{Z}}$ \\ \hline
 & & & \\
LEP parametrization & -0.44 $< \Delta\kappa_{\gamma} <$ 0.55 & -0.10 $< \lambda <$ 0.11 & -0.12 $< \Delta g_{1}^{Z} <$ 0.20 \\
Equal couplings & -0.16 $< \Delta\kappa <$ 0.23 & -0.11 $< \lambda <$ 0.11 &  \\
      \end{tabular}
    \end{ruledtabular}
 \end{table*}

 \begin{table*}[htb]
    \caption{
   Comparison of 95\% C.L. one-parameter TGC limits between the different channels 
   studied at D0 with $\approx{1}~\rm{fb^{-1}}$ of data: $WW\rightarrow \ell\nu \ell\nu$,
   $W\gamma\rightarrow \ell\nu\gamma$, $WZ\rightarrow \ell\ell\ell\nu$ and
   $WW+WZ\rightarrow \ell\nu{jj}$ ($l=\mu,e$) at $\Lambda_{NP}=$
   2~TeV.}
    \label{tab:compareLimits}
    \begin{ruledtabular}
      \begin{tabular}{l c c c}
LEP parametrization & $\Delta\kappa_{\gamma}$ & $\lambda=\lambda_{\gamma}=\lambda_{Z}$ & $\Delta{g_{1}^{Z}}$ \\ \hline\\
$WZ\rightarrow \ell\nu \ell\ell$ ($1~\rm{fb^{-1}}$) & - & -0.17 $<\lambda <$ 0.21 & -0.14 $<\Delta{g_{1}^{Z}} <$ 0.34 \\
$W\gamma\rightarrow \ell\nu\gamma$ ($0.7~\rm{fb^{-1}}$) & -0.51 $<\Delta\kappa_{\gamma} <$ 0.51 & -0.12 $<\lambda <$ 0.13 &  \\ 
$WW\rightarrow \ell\nu \ell\nu$ ($1~\rm{fb^{-1}}$) & -0.54 $<\Delta\kappa_{\gamma} <$ 0.83 & -0.14 $<\lambda <$ 0.18 & -0.14 $<\Delta{g_{1}^{Z}} <$ 0.30 \\ 
  $WW+WZ\rightarrow \ell \nu jj$ ($1.1~\rm{fb^{-1}}$) & -0.44 $<\Delta\kappa_{\gamma} <$ 0.55 & -0.10 $<\lambda <$ 0.11 & -0.12 $<\Delta{g_{1}^{Z}} <$ 0.20 \\ 
\hline\\
equal couplings & $\Delta\kappa_{\gamma}$ & $\lambda=\lambda_{\gamma}=\lambda_{Z}$ & $\Delta{g_{1}^{Z}}$ \\ \hline\\
$WZ\rightarrow \ell\nu \ell\ell$ ($1~\rm{fb^{-1}}$) &  & -0.17 $<\lambda <$ 0.21 & \\ 
  $W\gamma\rightarrow \ell\nu\gamma$ ($0.7~\rm{fb^{-1}}$) &  & -0.12 $<\lambda <$ 0.13  & \\
  $WW\rightarrow \ell\nu \ell\nu$ ($1~\rm{fb^{-1}}$) & -0.12 $<\Delta\kappa <$ 0.35 & -0.14 $<\lambda <$ 0.18 & \\ 
  $WW+WZ\rightarrow \ell\nu jj$ ($1.1~\rm{fb^{-1}}$) & -0.16 $<\Delta\kappa <$ 0.23 & -0.11 $<\lambda <$ 0.11&  \\ 
      \end{tabular}
    \end{ruledtabular}
  \end{table*}
  \begin{table*}[h]
    \caption{
    Measured values of $\kappa_{\gamma}$, $\lambda$ and $g_{1}^{Z}$ couplings and their associated uncertainties at 68\% C.L. 
    obtained from the one-parameter fits combining data from different topologies and energies at LEP2 experiments. 
    The last column shows the D0 result obtained from the $\ell\nu{jj}$ final states only selected from $1~\rm{fb^{-1}}$ of data. The
    uncertainties include both statistical and systematic sources.}
    \label{tab:singleLEP}
    \begin{ruledtabular}
      \begin{tabular}{l c c c c}
      68\% C.L. & ALEPH & OPAL & L3 & D0 ($\ell\nu{jj}$) \\ \hline\\
$\kappa_{\gamma}$ & 0.971$\pm$0.063  & 0.88$^{+0.09}_{-0.08}$     & 1.013$\pm$0.071  & 1.07$^{+0.26}_{-0.29}$ \\
$\lambda$         & -0.012$\pm$0.029 & -0.060$^{+0.034}_{-0.033}$ & -0.021$\pm$0.039 & 0.00$^{+0.06}_{-0.06}$ \\
$g_{1}^{Z}$       & 1.001$\pm$0.030  & 0.987$^{+0.034}_{-0.033}$  & 0.966$\pm$0.036  & 1.04$^{+0.09}_{-0.09}$ \\
      \end{tabular}
    \end{ruledtabular}
  \end{table*}

  As shown in Table~\ref{tab:compareLimits}, the 95\% C.L. limits on
  anomalous couplings $\Delta\kappa_{\gamma},~\Delta\lambda$, and
  $\Delta{g_{1}^{Z}}$ set using the dijet $p_{T}$ distribution of
  $WW/WZ\rightarrow \ell\nu jj$ events are comparable with the 95\%
  C.L.  limits set by the D0 Collaboration from $WW$~\cite{bib:ww},
  $WZ$~\cite{bib:wz}, and $W\gamma$~\cite{bib:wgamma} production in
  fully leptonic channels using $\approx{1}~\rm{fb^{-1}}$ of data.
  The most recent 95\% C.L. one-parameter limits from the CDF
  Collaboration under the equal couplings scenario at
  $\Lambda_{NP}=1.5$~TeV are $-0.46<\Delta\kappa <0.39$ and
  $-0.18<\lambda <0.17$ using 350 pb$^{-1}$ of data, combining the
  $\ell\nu{jj}$ and $\ell\nu\gamma$ ($l=e,\mu$) final
  states~\cite{bib:cdf}.  These results are limited by statistics, but a
  factor of nearly 10 times more data is expected to be available for
  analysis by D0 by the end of Run II of the Fermilab Tevatron.  With
  additional data the potential to reach the individual LEP2 anomalous
  TGC limits~\cite{bib:aleph,bib:opal,bib:l3} shown in
  Table~\ref{tab:singleLEP} is significant.  
  The combined LEP2 results still represent the world's tightest limits 
  on charged anomalous couplings~\cite{bib:anom-lep1} and give the most 
  probable values of $\kappa_{\gamma},~\lambda$, and $g_{1}^{Z}$ as
  $\kappa_{\gamma}=0.973^{+0.044}_{-0.045}$, $\lambda
  =-0.028^{+0.020}_{-0.021}$, and $g_{1}^{Z}=0.984^{+0.022}_{-0.019}$
  at 68\% C.L.

  In summary, we have presented a measurement of $\gamma{WW}/ZWW$
  couplings using a sample of semileptonic decays of $WW/WZ$ boson pairs 
  corresponding to 1.1~\ifb\ of \pp\ collisions collected with the D0 
  detector at the Fermilab Tevatron Collider.  The measurement is in 
  agreement with the SM.  On the other hand, this analysis yields the most 
  stringent limits on $\gamma{WW}/ZWW$ anomalous couplings from the 
  Tevatron to date, complementing similar measurements performed in 
  fully leptonic decay modes from $W\gamma$, $WW$, and $WZ$ production.

  \begin{figure*}[tbp] 
    \begin{centering}
        \includegraphics[width=8.8cm]{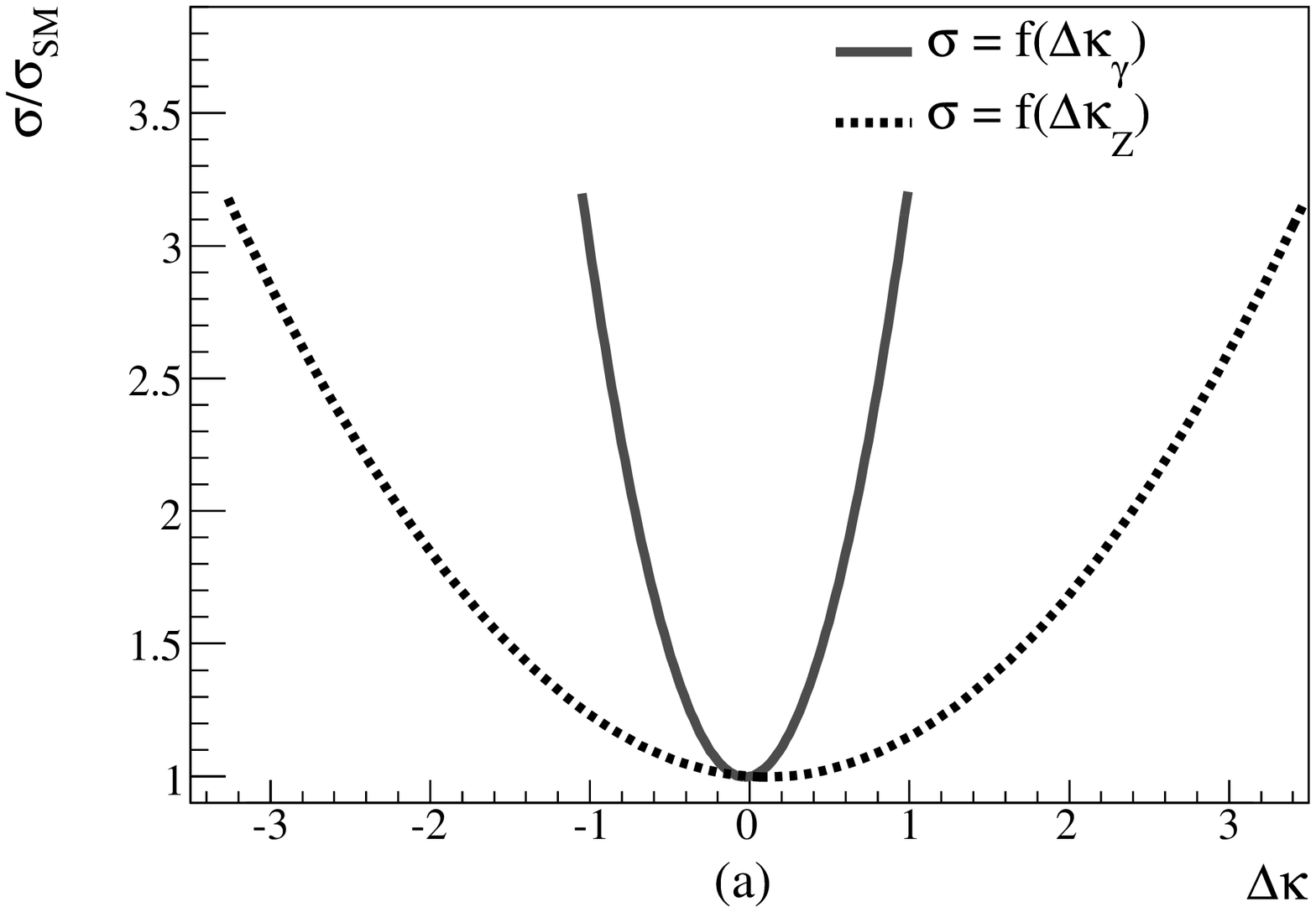}
        \includegraphics[width=8.8cm]{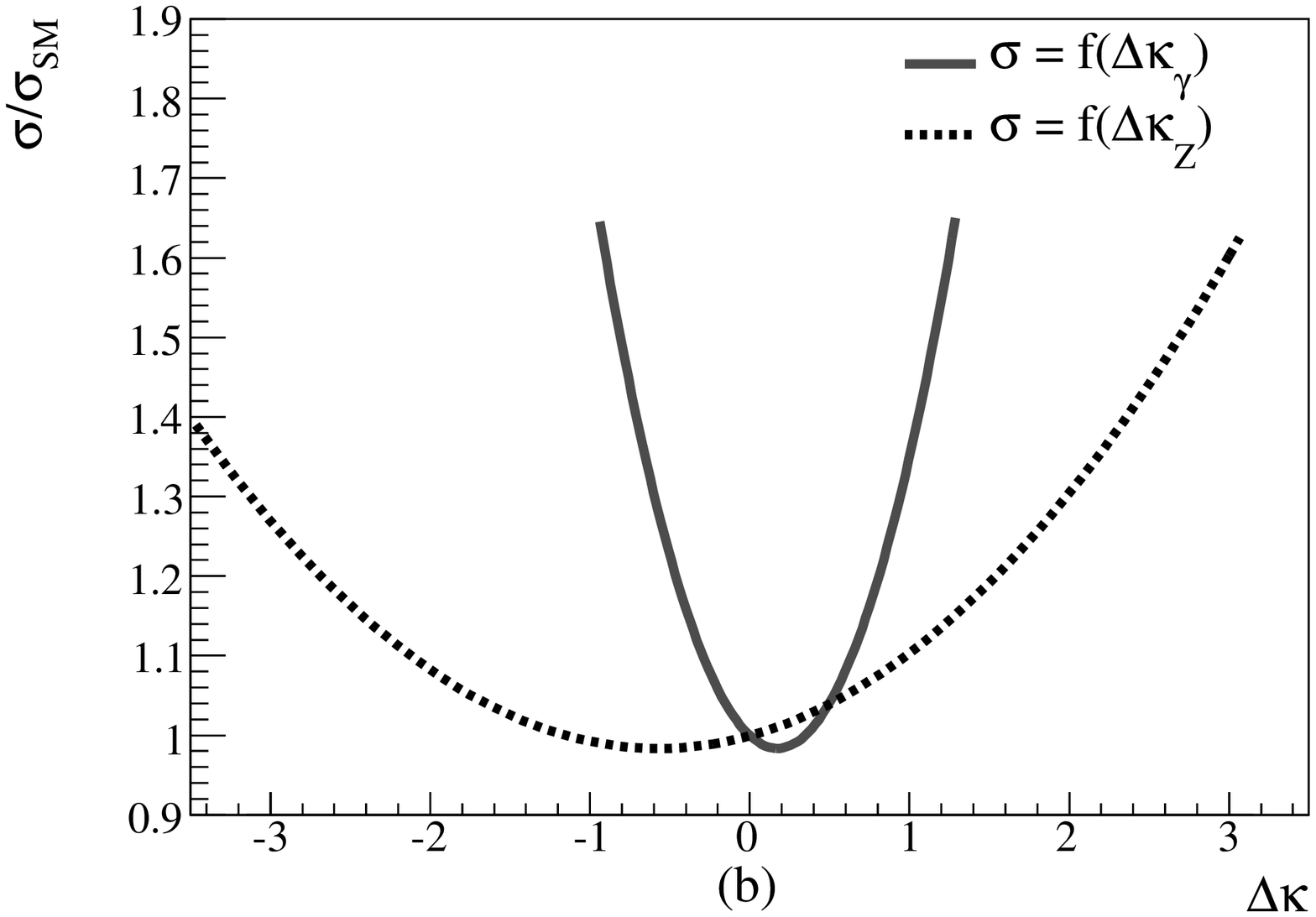}\\

      \caption{ Semileptonic production cross sections for (a) $WW$
      and (b) $WZ$ normalized to the SM prediction as a function of
      anomalous coupling $\Delta\kappa$ ($\lambda =
      \Delta{g_{1}^{Z}}=0$) in the LEP parametrization
      scenario. The new physics scale $\Lambda_{NP}$ is set to 2~TeV.}
      
      \label{fig:x-secs}
    \end{centering}
  \end{figure*}

  \begin{figure*}[tbp] 
    \begin{centering}
        \includegraphics[width=8.8cm]{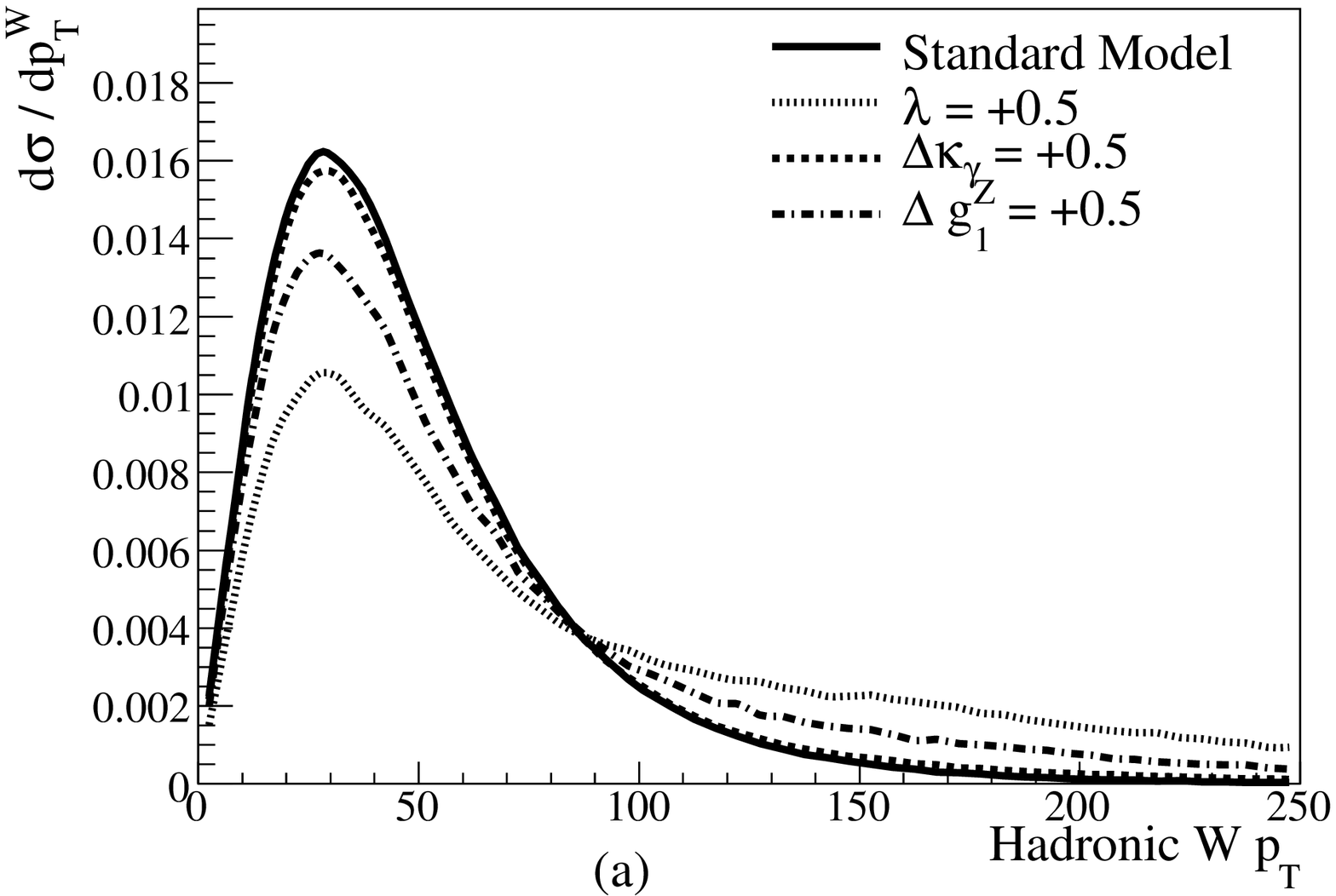}
        \includegraphics[width=8.8cm]{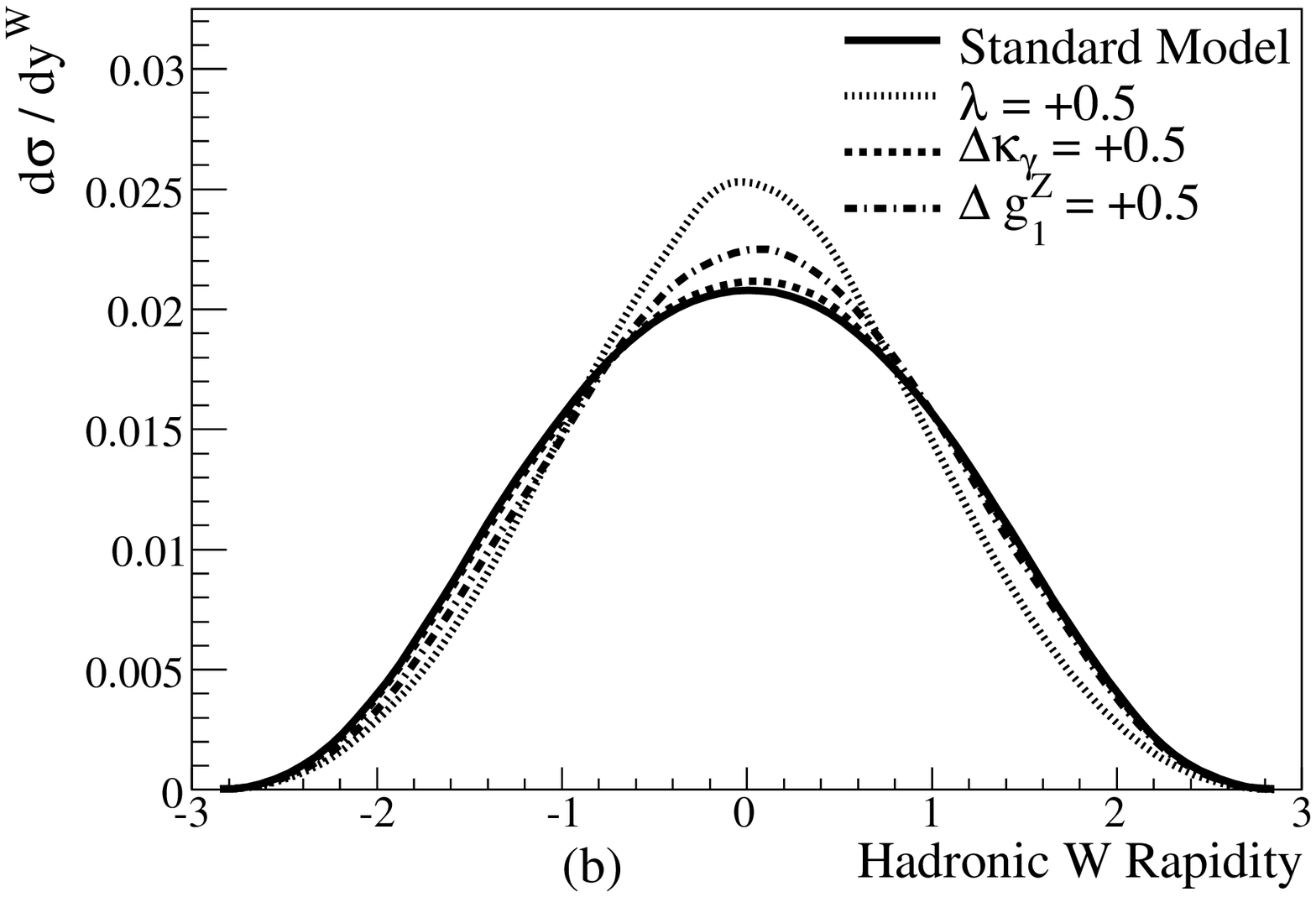} \\

      \caption{ Normalized distributions of the hadronic W boson (a)
$p_T$ and (b) rapidity at the parton level in $WW$ production
including anomalous couplings under the LEP parametrization
scenario: $\Delta\kappa_{\gamma}=$ +0.5 ($\lambda = \Delta{g_{1}^{Z}}=$ 0,
      $\Delta\kappa_{Z}=$ -0.15), $\lambda=$ +0.5
      ($\Delta\kappa_{\gamma} = \Delta\kappa_{Z} = \Delta{g_{1}^{Z}}=$
      0), and $\Delta g_{1}^{Z}=$ +0.5 ($\Delta\kappa_{\gamma}=\lambda
      =$ 0, $\Delta\kappa_{Z}=$ 1.5) compared to the SM distribution
      for $WW$ production with unity normalization. The new physics
      scale $\Lambda_{NP}$ is set to 2~TeV.}

      \label{fig:kine}
    \end{centering}
  \end{figure*}

   \begin{figure*}[htb] \begin{centering} 
     \includegraphics[width=8.8cm]{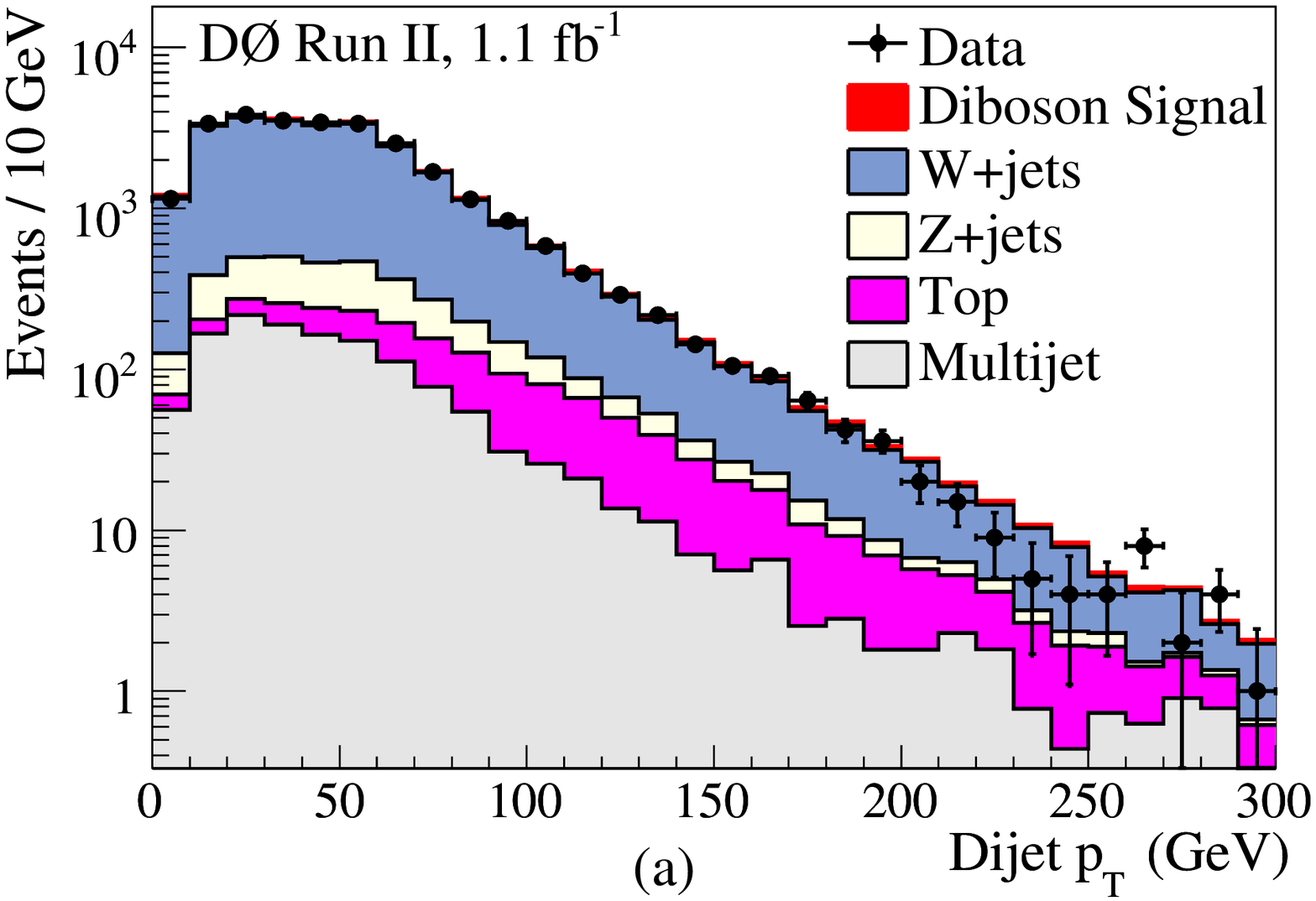}
     \includegraphics[width=8.8cm]{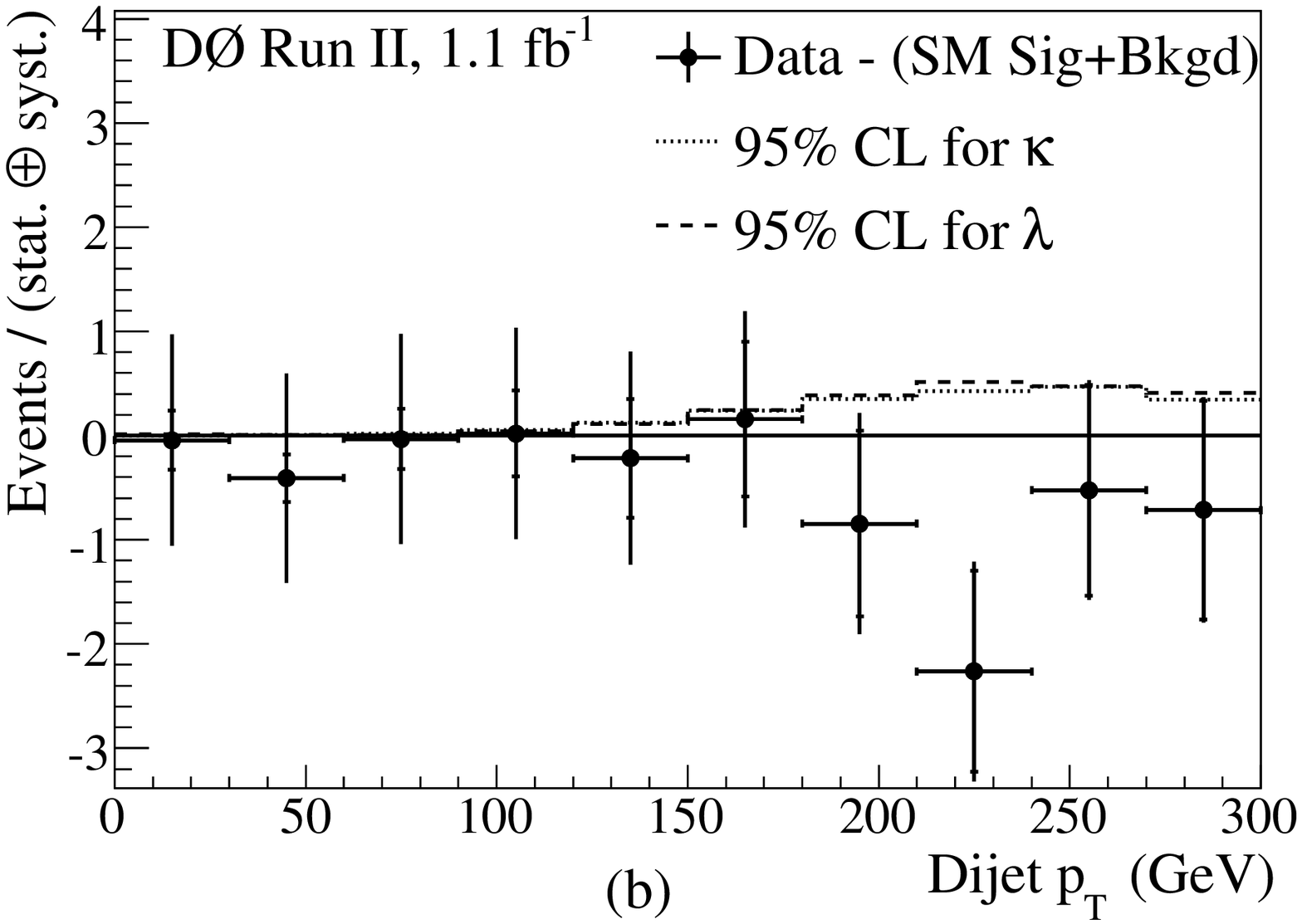}\\
     
     \caption{(a) The dijet $p_{T}$ distribution of combined
     (electron+muon) channels for data and SM predictions following
     the fit of MC to data. (b) The difference between data and
     simulation divided by the uncertainty (statistical and
     systematic) for the dijet $p_{T}$ distribution.  Also shown are
     the MC signals for anomalous couplings corresponding to the 95\%
     C.L. limits for $\Delta\kappa$ and $\lambda$ in the LEP
     parametrization scenario.  The full error bars on the data
     points reflect the total (statistical and systematic)
     uncertainty, with the ticks indicating the contribution due only
     to the statistical uncertainty. }
  
    \label{fig:subtracted} 
    \end{centering} 
  \end{figure*}

  \begin{figure*}[htb] 
    \begin{centering}
      \includegraphics[width=8.8cm]{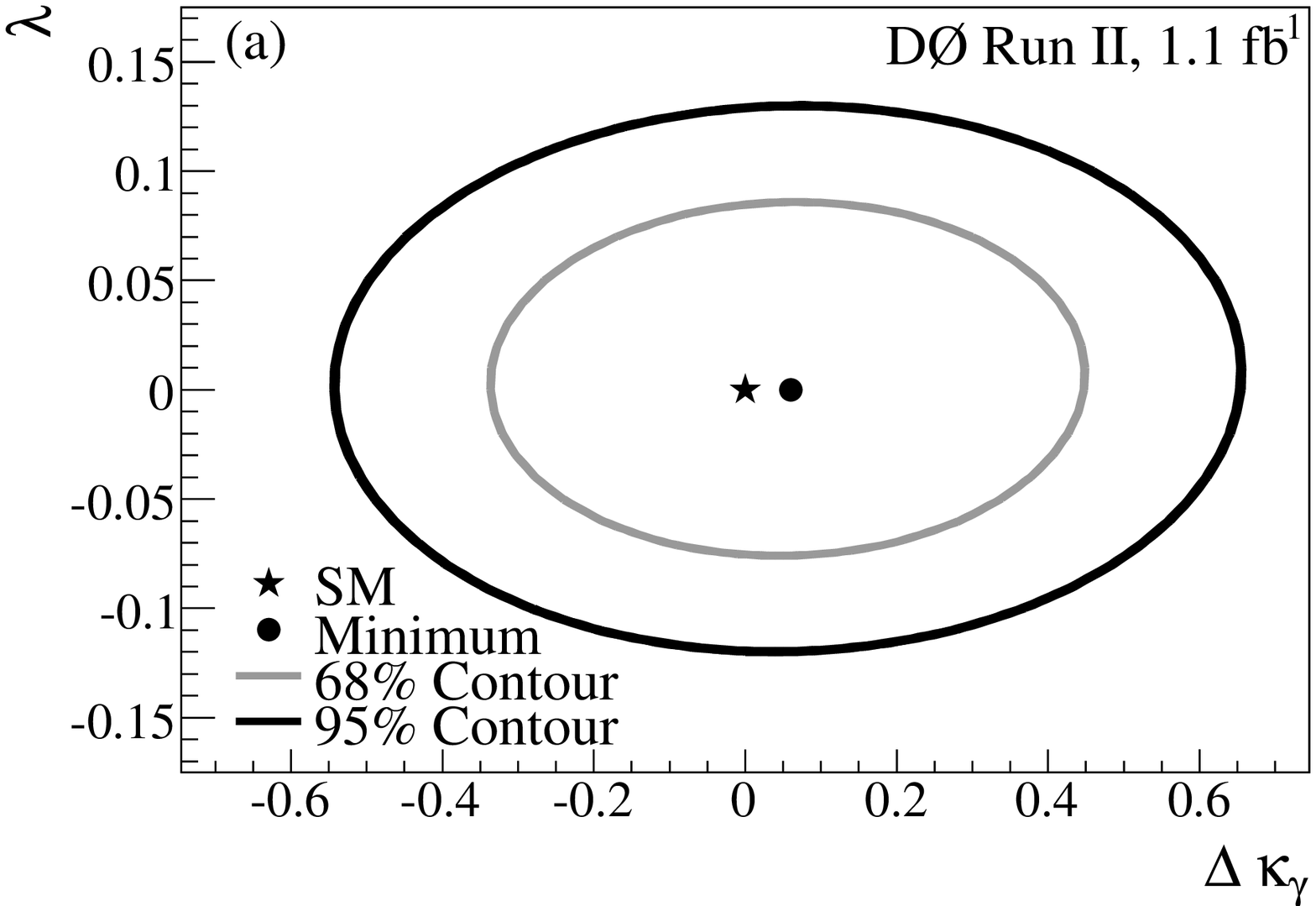}
      \includegraphics[width=8.8cm]{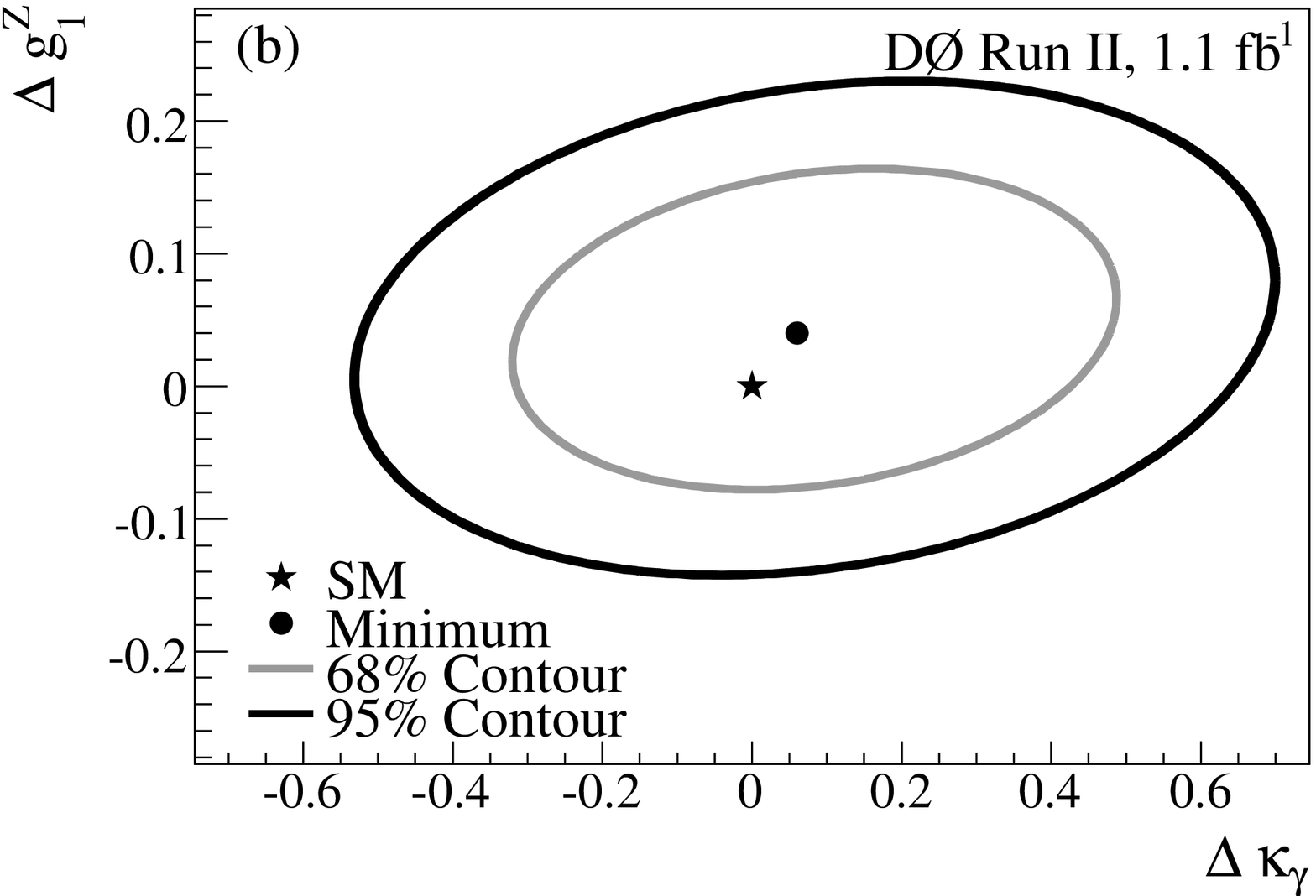} \\
      \includegraphics[width=8.8cm]{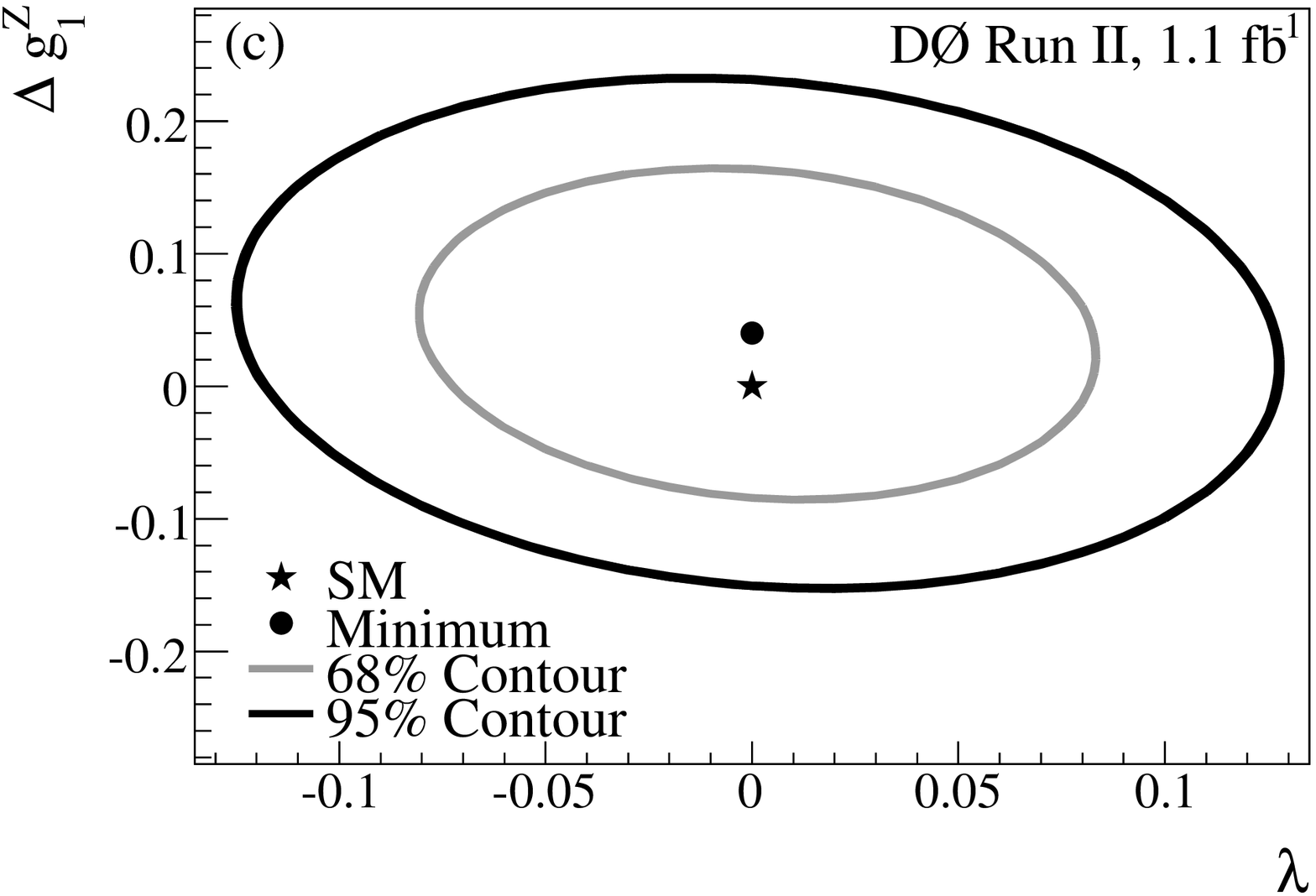}
  \caption{The 68\% C.L. and 95\% C.L. two-parameter limits on the $\gamma WW/ZWW$ 
  coupling parameters $\Delta\kappa_{\gamma}$, $\lambda$, and $\Delta g_{1}^{Z}$, in
  the LEP parametrization scenario and $\Lambda_{NP}=$ 2~TeV. The dots indicate the most probable 
  values of anomalous couplings from the two-parameter combined (electron+muon) fit and the star markers denote the SM prediction. 
  \label{fig:limit1}} \end{centering} \end{figure*}
  \begin{figure*}[htb] 
    \begin{centering}
      \includegraphics[width=8.8cm]{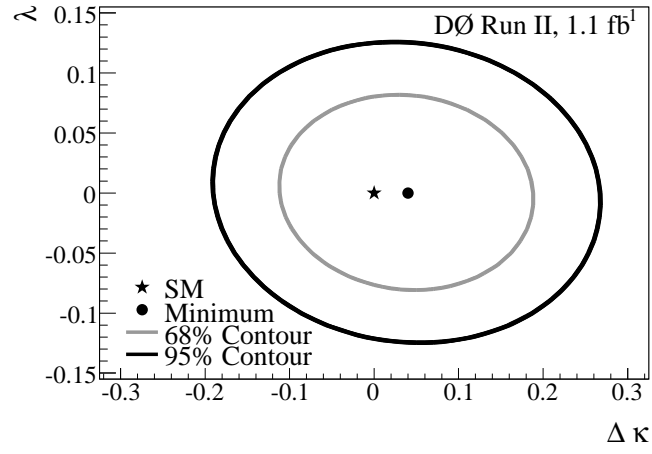}
  \caption{The 68\% C.L. and 95\% C.L. two-parameter limits on the $\gamma WW/ZWW$ 
  coupling parameters $\Delta\kappa$ and $\lambda$, in the equal couplings scenario 
  and $\Lambda_{NP}=$ 2~TeV. The dot indicates the most probable 
  values of anomalous couplings from the two-parameter combined (electron+muon) fit and the star marker denotes the SM prediction.
  \label{fig:limit2}} \end{centering} \end{figure*}

  \clearpage
  \input acknowledgement_paragraph_r2.tex   % input acknowledgement

\end{document}
%
% ****** End of file template.aps ******

%% file: list_of_authors_r2.tex
% LIST_OF_AUTHORS_R2.TEX                 7/7/09             
%
\author{V.M.~Abazov$^{37}$}
\author{B.~Abbott$^{75}$}
\author{M.~Abolins$^{65}$}
\author{B.S.~Acharya$^{30}$}
\author{M.~Adams$^{51}$}
\author{T.~Adams$^{49}$}
\author{E.~Aguilo$^{6}$}
\author{M.~Ahsan$^{59}$}
\author{G.D.~Alexeev$^{37}$}
\author{G.~Alkhazov$^{41}$}
% ...

\affiliation{\vspace{0.1 in}(The D\O\ Collaboration)\vspace{0.1 in}}
% ...
\affiliation{$^{6}$University of Alberta, Edmonton, Alberta, Canada;
                Simon Fraser University, Burnaby, British Columbia, Canada;
                York University, Toronto, Ontario, Canada and
                McGill University, Montreal, Quebec, Canada}
% ...
\affiliation{$^{30}$Tata Institute of Fundamental Research, Mumbai, India}
% ...
\affiliation{$^{37}$Joint Institute for Nuclear Research, Dubna, Russia}
% ...
\affiliation{$^{41}$Petersburg Nuclear Physics Institute,
                St. Petersburg, Russia}
% ...
\affiliation{$^{49}$Florida State University, Tallahassee, Florida 32306, USA}
% ...
\affiliation{$^{51}$University of Illinois at Chicago,
                Chicago, Illinois 60607, USA}
% ...
\affiliation{$^{59}$Kansas State University, Manhattan, Kansas 66506, USA}
% ...
\affiliation{$^{65}$Michigan State University,
                East Lansing, Michigan 48824, USA}
% ...
\affiliation{$^{75}$University of Oklahoma, Norman, Oklahoma 73019, USA}
% ...

%% file: acknowledgement_paragraph_r2.tex
% acknowledgement_paragraph_r2.tex                         7/7/09
%
We thank the staffs at Fermilab and collaborating institutions, 
and acknowledge support from the 
DOE and NSF (USA);
CEA and CNRS/IN2P3 (France);
FASI, Rosatom and RFBR (Russia);
CNPq, FAPERJ, FAPESP and FUNDUNESP (Brazil);
DAE and DST (India);
Colciencias (Colombia);
CONACyT (Mexico);
KRF and KOSEF (Korea);
CONICET and UBACyT (Argentina);
FOM (The Netherlands);
STFC and the Royal Society (United Kingdom);
MSMT and GACR (Czech Republic);
CRC Program, CFI, NSERC and WestGrid Project (Canada);
BMBF and DFG (Germany);
SFI (Ireland);
The Swedish Research Council (Sweden);
CAS and CNSF (China);
and the
Alexander von Humboldt Foundation (Germany).

%% file: prd.bbl
\begin{thebibliography}{99}
\bibitem{bib:footnote} D0 uses a cylindrical coordinate system with the $z$ axis running along the beam axis. Angles $\theta$ and $\phi$ are the polar and azimuthal angles, respectively. Pseudorapidity is defined as $\eta = -\ln \left[\tan(\theta/2) \right]$, in which $\theta$ is measured with respect to the proton beam direction. In the massless limit, $\eta$ is equivalent to the rapidity $y = (1/2) \ln \left[ (E+p_z)/(E-p_z)\right]$. $\eta_{\rm det}$ is the pseudorapidity measured with respect to the center of the detector.
\end{thebibliography}
